\documentclass[aps,pra,twocolumn]{revtex4-1}
\usepackage{epsfig}
\usepackage{amsbsy,latexsym}
\usepackage{amsmath}
\usepackage{amssymb, mathrsfs}
\usepackage[mathscr]{eucal}
\usepackage{hyperref}
\newcommand{\Rangle}{\rangle\!\rangle}
\newcommand{\Langle}{\langle\!\langle}
\newcommand{\hilb}{\mathcal{H}}
\def\Tr{{\rm Tr}}
\def\map#1{{\mathcal #1}}
\def\>{\rangle}
\def\<{\langle}
\def\qed{$\blacksquare$}
\def\Lin{{\mathsf{Lin}}}
\newtheorem{lemma}{Lemma}
\newtheorem{theo}{Theorem}
\newtheorem{prop}{Proposition}
\begin{document}
\title{Optimal Probabilistic Simulation of Quantum Channels from the Future to
the Past}

\author{Dina Genkina} 
\affiliation{Perimeter Institute for Theoretical Physics, 31 Caroline Street
North, Waterloo, Ontario N2L 2Y5, Canada.} 

\author{Giulio Chiribella} 
\affiliation{Perimeter Institute for Theoretical Physics, 31 Caroline Street
North, Waterloo, Ontario N2L 2Y5, Canada.} 

\author{Lucien Hardy} 
\affiliation{Perimeter Institute for Theoretical Physics, 31 Caroline Street
North, Waterloo, Ontario N2L 2Y5, Canada.} 

\date{\today}

\begin{abstract}
 We introduce the study of quantum protocols that probabilistically simulate
quantum channels from a sender in the future to a receiver in the past. 
 The maximum probability of simulation is determined by causality and depends on
the amount and type (classical or quantum) of information that the channel can
transmit. We illustrate this dependence in several examples, including ideal
classical and quantum channels, measure-and-prepare channels, partial trace
channels, and universal cloning channels.   For the simulation of partial trace
channels, we consider generalized teleportation protocols that take $N$ input
copies of a pure state in the future and produce $M\le N$ output copies of the
same state in the past. In this case, we show that the maximum probability of
successful teleportation increases with the number of input copies,  a feature
that was impossible in classical physics.  In the limit of asymptotically large
$N$, the probability converges to the probability of simulation for an ideal
classical channel.    
 Similar results are found for universal cloning channels from $N$ copies to $M 
> N$ approximate copies, exploiting a time-reversal duality between universal
cloning and partial trace.       
\end{abstract}

\maketitle
\section{Introduction}
Quantum Theory is generally formulated relative to a given causal structure.
This is the case both in Quantum Field Theory, where the spacetime metric is
given from the beginning, and in the operational framework of Quantum
Information, where protocols and computations consist in  sequences of
operations performed at different times.   Relative to the given causal
structure, Quantum Theory has to satisfy the \emph{causality principle}
\cite{puri,QMfrompuri}, stating that the probability of a measurement outcome at
a given time be independent of the choice of operations performed at later
times.

\begin{figure}[h]
\begin{centering}
\includegraphics[scale=0.3]{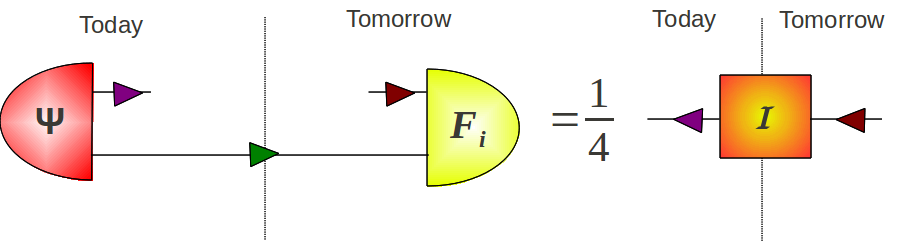}
\par \end{centering}  
\caption{\label{fig:originaltele} (Color online) Probablilistic teleportation
from the future
to
the past. Today Alice prepares    a maximally entangled two-qubit state $\Psi$
of the ancilla and the output system, while tomorrow Bob performs a measurement
on the
ancilla and his input qubit. The measurement has four possible
outcomes, for one of which (here denoted by $i_0$) successful teleportation will
occur. With
probability $p= \frac{1}{4}$ Alice obtains Bob's input state today. However, she
will not know whether  teleportation has been successful until Bob performs his
measurement and communicates the outcome.}
\end{figure}

The causality principle forbids any form of signalling from the future to the
past:  A sender  in the future cannot deterministically transfer the state of
his
system to a receiver  in the past.  However, it is easy to imagine situations
where the state of a system is transferred from the future to the past 
\emph{with some probability}, without leading to signalling.   This is what
happens, for example, if we eliminate the communication of classical data in the
original quantum teleportation protocol \cite{Original}:  The receiver in the
past  (say, Alice) can  prepare a maximally entangled state and the sender in
the future  (say, Bob)  can perform a Bell measurement on the state to be
teleported together with half of the entangled state, as in Fig. 
\ref{fig:originaltele}.
 In this way, upon observing the right measurement outcome, Bob will
know that
the state has been teleported successfully to the past, and Alice will not need
to perform  any correction operation on the output.  
Of course, to know that teleportation has succeeded, Bob  needs  to wait until
he sees the  outcome of the Bell measurement.
However, the important fact here is that, even before the measurement is
performed, one half of the entangled state is already ready to be used by Alice
in a quantum circuit:  Alice  can perform any desired computation on her system,
and in the end, if teleportation is successful, Bob will know that Alice's
computation has been performed on the input state he provided in the future.  
Conditionally to the occurrence of the right outcome, everything behaves as
though the input state has travelled from the future to the past through an
ideal quantum channel.   Note that this fact is not in contradiction with
causality:  The probability of teleportation is small enough that no signal can
be sent from Bob to Alice.    We will refer to the above use of probabilistic
teleportation as a \emph{probabilistic simulation of the identity channel from
the future to the past}.   

The idea of probabilistic teleportation as a simulated time-travel was
originally proposed by Bennett and Schumacher in an unpublished work
\cite{bennett's-talk}, and, more recently, by Svetlichny \cite{svetlichny}.   
In different terms, this idea also appeared implicitly in the context of the
graphical language of categorical quantum mechanics \cite{abracoecke,coecke},
where the reinterpretation of probabilistic teleportation as information flow
from the future to the past follows from a stretching of wires in the basic
teleportation diagram.    Interestingly, similar ideas also appeared in the work
of Horowitz and Maldacena \cite{horocena} in the context of black hole
evaporation, as an attempt to reconcile the unitarity of the black hole
$S$-matrix  with Hawking's semiclassical arguments (see also the discussion by
Gottesmann and Preskill \cite{preskgott}).  Recently, the role of probabilistic
teleportation in simulating  closed
timelike curves has been further explored by Lloyd \emph{et al} in Refs.
\cite{CTCs,
CTCs2}. In particular, Ref. \cite{CTCs} reports an experiment that uses
probabilistic teleportation to simulate a quantum computation within a closed
timelike curve.   De Silva, Galvao and Kashefi \cite{galvao} showed that some
patterns in measurement-based quantum computation can be interpreted as
deterministic simulations of  closed timelike curves.

All the works mentioned so far focused on the use of teleportation for the
probabilistic simulation of an ideal quantum channel from the future to the
past.  However, there are many interesting scenarios where one needs to consider
the simulation of more general quantum channels.  For example, suppose that we
have $N$ identical copies of a the same state and that we want to teleport just
one copy to the past.  Does the probability of success increase with the number
if input copies?  And, if it does, what is the the asymptotic value of the
success probability in the limit  $N \to \infty$?   To answer these questions we
have to address the probabilistic simulation of channels that trace all  systems
but one.

\begin{figure}[h]
\begin{centering}
\includegraphics[scale=0.3]{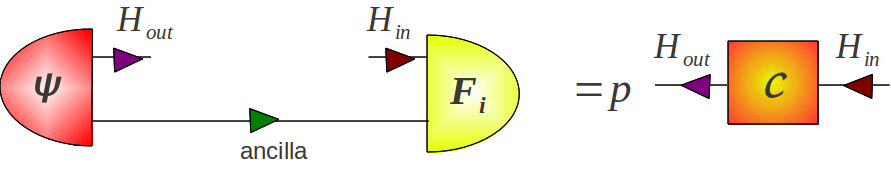}
\par \end{centering}

\caption{\label{genchan}  
(Color online) Probabilistic simulation of a given quantum channel from the
future to the past
via a generalized teleportation scheme. The output system and an ancilla are
first  prepared in a bipartite state $\Psi$. When the input becomes available,
the input state and the ancilla are measured jointly.   For a particular
measurement outcome $i_0$ the simulation will be successful:  The effective
transformation from input to output will be proportional to the desired quantum
channel, with proportionality constant $p$ equal to the probability of
successful simulation.} 
\end{figure}

 In this paper we will address the general problem of the probabilistic
simulation of a given channel from the future to the past, showing how the
causality principle determines the  maximum probability of success. To find the
maximum probability we will optimize over all generalized teleportation schemes
where a bipartite state is prepared and the input of the channel is jointly
measured along with half of the bipartite state, so that, for a particular
outcome, the desired channel is simulated, as in Fig. \ref{genchan}.  
The spirit of this work is similar to the spirit of the early works on the
optimal
cloning of non-orthogonal quantum states
\cite{hillbuz,gismass,cloning,contvar,phase}:   In that case, one knew from the
no-cloning theorem \cite{no-cloning,no-cloning2} that it is impossible to
produce perfect copies of the input
state   and the goal was to find the optimal physical process that approximates
the impossible cloning transformation.   In our case, we know from causality
that it is impossible to have a channel that deterministically transfers
information from the future to the past and our goal is to find the optimal
process that achieves the desired channel with maximum probability of success. 
In the same way in which the study of optimal cloning shed light on the process
of copying information in the quantum world, we expect that the study of optimal
simulations of channels to the past  will shed light the interplay between
causal structure and quantum information flow.

The main message of this paper is that  the maximum probability of simulating a
quantum channel from the future to the past is a decreasing function of the
amount of information that the channel can transmit.
This general feature   will be illustrated in several examples.  First, we will
consider ideal
classical channels (those that perfectly transmit the states of an orthonormal
basis)
and, second,  ideal quantum channels (those that perfectly transmit all states
of a given
quantum system).   In the first case we find the maximum probability 
\begin{align}
p_{cl}  =  \frac 1 d, 
\end{align}
where $d$ is the dimension of the Hilbert space, while in the second case we
find  
\begin{align}
p_q  =  \frac 1 {d^2},
\end{align}  
which is exactly the probability of the outcome that does not require correction
operations in the original
teleportation protocol \cite{Original}.    
We will then focus on the probabilistic simulation of measure-and-prepare
channels, which are a class of channels that transmit only classical
information.   In this case, we will find that the maximum probability of
success is at least equal to the probability $p_{cl}  = 1/d$ for a classical
channel with the same output Hilbert space.   The exact value of the probability
of success
will be computed in two relevant cases of measure-and-prepare channels: the
channel associated to the optimal estimation of a pure state, and the universal
NOT channel \cite{UNOT}.   
Finally, we will analyze in detail the case of trace channels, sending $N$ 
copies of a given pure state to $M\le N$ copies of the same state, and the case
of universal cloning channels \cite{cloning}, sending $N$ copies of a pure state
to $M >N$
optimal approximate copies.  In both cases we find that the maximum probability
of success is given by 
 \begin{align}\label{genNMbound}
 p^+_{q, N \to M} =  \frac{d^{(|N-M|)}_+}{ d^{(N)}_+  d^{(M)}_+}  \qquad 
d^{(k)}_+  :=   
  \begin{pmatrix}    
  d+  k-1\\
  k
    \end{pmatrix}  , 
 \end{align}
where $d$ is the dimension of the one-particle Hilbert space and the superscript
in  $p^+_{q, N \to M}$ reminds that we are restricting ourselves to the
symmetric subspaces, in which the $N$ input copies and the $M$  output copies of
the given pure state live. 
 In the particular case of $M=1$ Eq. (\ref{genNMbound})  yields the value
\begin{align}
p^+_{q, N \to 1}  =  \frac N   {d(d+N-1)},
\end{align} 
which increases with the number of input copies, starting from $p^+_{q, 1\to1} 
=  1/d^2$ and reaching  the classical value $p_{cl}= 1/d$ in the asymptotic
limit $N \to \infty$.  This result has to be contrasted with the classical
scenario, where having more input copies of a pure state cannot lead to any
improvement:   Since classical pure states can be perfectly cloned, there is no
difference in having more input copies.   The convergence of the
probability of success to the classical value $p_{cl}  = 1/d$  will be explained
as a consequence of the convergence of the trace channel to a
measure-and-prepare channel \cite{definetti}.  
Concerning  the probability  $p_{q, N\to M}^+$  in Eq. (\ref{genNMbound}), it 
is also worth noting that it   is symmetric in $N$ and $M$:  In other words, the
probability of successful simulation for a trace channel from $N$ to $M \le  N$
copies is equal to the probability of successful simulation for a universal
cloning channel from $M$ to $N \ge M$  copies.  We will explain the symmetry of
the probabilities as a consequence of a \emph{time-reversal duality} between
trace and universal cloning.

As we anticipated, the message emerging from our results is that the maximum
probability for a given channel is a decreasing function of the amount of
information that the channel can transmit:  The smallest value $p_q=  1/d^2$
corresponds to the identity channel, while measure-and-prepare channels have
probability $p \ge 1/d$, and the erasure channels $\mathcal C (\rho)  = 
\Tr[\rho] ~ \rho_0$ have probability $p= 1$.   We will also make the connection
more quantitative, providing a set of  \emph{statistical information bounds}, 
stating
that  the probability of success in the simulation of a channel is upper bounded
by the inverse   of the amount of  information  that the channel can transmit. 
The
amount of information  will be quantified here as the maximum payoff that two
parties can achieve in a communication game.

 The structure of the paper is the following:  In Section  \ref{sec:general}  we
present the general method and, in particular, the \emph{causality bound}, a
necessary and sufficient condition for the existence of a probabilistic
simulation with probability $p$.  Then, we will analyze the probabilistic
simulation  of several channels:  Sections \ref{sec:ideal}, \ref{sec:measprep},
and \ref{sec:opttele}    will focus on ideal classical and quantum channels,  on
measure-and-prepare channels, and on partial trace channels from $N$ to $M \le
N$ 
copies, respectively.   Section \ref{sec:opttele} will also discuss the case of
universal cloning channels from $N$ to $M> N$ copies, exploiting the existence
of a time-reversal duality between partial trace channels and cloning channels. 
Finally, in Section \ref{sec:summary} we will derive the statistical information
bound and use it to explain the asymptotic behaviour of the maximum probability
for partial  trace and cloning channels.   The conclusions of the paper are
drawn in
Section \ref{sec:conclusion}.

\section{Theoretical framework}\label{sec:general}
To find the maximum probability to simulate a given channel we will use
formalism of the Choi
operators \cite{choi} and the framework of quantum combs
\cite{prlcombs,pracombs}.   In this
section we present the basic notions used in the paper and we derive the
\emph{causality bound}, a necessary and sufficient condition for the simulation
of a given channel with probability $p$.

\subsection{Causality bound for quantum operations in a generalized
teleportation protocol}
\begin{figure}[h]
\begin{centering}
\includegraphics[scale=0.3]{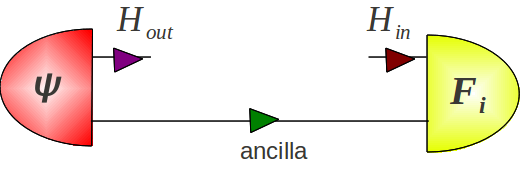}
\par \end{centering}
\caption{\label{gentele} (Color online) Generalized teleportation circuit
consisting of a
joint state preparation and of a joint measurement.}
\end{figure}

Any quantum circuit can be represented as a collection of preparations,
measurements, transformations, and quantum wires along which states travel
undisturbed. In this paper we only consider generalized teleportation  circuits,
which
consist of one joint preparation of output and ancilla and one joint measurement
on the
input and ancilla,  as in the circuit of Fig.  \ref{gentele}. 

Here the preparation is represented by  a joint quantum state  $|\Psi\>  \in 
\hilb_A  \otimes  \hilb_{out}  $ of  the ancilla and the output system, and  the
quantum measurement is represented by   a \emph{positive operator
valued measure (POVM)}   $\{F_i\}_{i \in X}$,  namely a collection of positive
operators $F_i \ge 0 $ acting on the Hilbert space  $\hilb _{in} \otimes
\hilb_{A}$ with the normalization property   
\begin{align}\label{eq:alloraaa}
\sum_{i \in X} F_i = I_{in} \otimes  I_{A} ,  
\end{align}
$I_{in}$ and $I_A$ representing the identity operators on the input Hilbert
space $\hilb_{in}$  and on the ancilla Hilbert space $\hilb_A$, respectively.

Each possible outcome  ${i\in X}$ corresponds to a different linear map
$\mathcal{E}_{i}$, mapping
density matrices on the input Hilbert space $\mathcal{H}_{in}$ to density
matrices on the output Hilbert space $\mathcal{H}_{out}$, and given by 
\begin{align*}
\map E_i (\rho)  :=         \Tr_{in, A}  [    (F_i  \otimes I_{out})   ( \rho 
\otimes   |\Psi\>\<\Psi|  ) ] ,  
\end{align*}   
$\Tr_{in,A}$ denoting the partial trace over the Hilbert spaces $\hilb_{in}$ and
$\hilb_{A}$.    
By construction, each linear map $\map E_i$  is completely positive and trace
non-increasing, namely it is a \emph{quantum operation} \cite{kraus}.  Note
that, however, in this case the output of the quantum operation precedes the
input in time, differing to the standard case originally considered by Kraus
\cite{kraus}.

Consider now the sum over all possible maps  $\{\map E_i\}_{i \in  X }$
corresponding to different measurement outcomes.  Summing over all possible
measurement outcomes is equivalent to  ignoring the measurement  outcome,  thus
obtaining the completely positive trace-preserving map (also known as 
\emph{quantum
channel}) $\map E_X$ given by
\begin{align*}
\map E_X  (\rho)  &= \sum_{i \in X}    \map E_i  (\rho)  \\
  & =  \sum_{i \in X}       \Tr_{in, A}  [    (F_i  \otimes I_{out})   ( \rho 
\otimes   |\Psi\>\<\Psi|  ) ]  \\
  &=    \sum_{i \in X}       \Tr_{in, A}  [     \rho  \otimes   |\Psi\>\<\Psi|  
] \\
  &=  \rho_0 ~  \Tr_{in}[\rho]   \qquad \rho_0  :=  \Tr_A[  |\Psi\>\<\Psi|],
\end{align*} 
the third equality coming from the normalization of Eq. (\ref{eq:alloraaa}).
Note that the trace-preserving map $\map E_X$ is an \emph{erasure channel}:  It
always produces the same output state $\rho_0$ independently of the input.  
 The fact that $\map E_X$ is an erasure channel is a manifestation of the
causality principle stating that no signal can be sent from the future back to
the past.  In the following we will use the notation $\map E_X =  \rho_0  ~
\Tr_{in}$ to mean that $\map E_X$ is the map defined by $\map E_X (\rho)  = 
\rho_0  \Tr_{in} [\rho] $, for every quantum state $\rho$ on $\hilb_{in}$.    
 
Since  the linear maps $\{\map E_i\}_{i \in X}$  are all completely positive, 
we must have   
 \begin{align}\label{causbound}
 \mathcal{E}_{i}\leq  \rho_0  ~ \Tr_{in}  \qquad \forall i \in X, 
 \end{align} 
where the inequality $\map A  \le \map B$  for two linear maps $\map A$ and
$\map B$ means that the linear map $\map B -  \map A$ is  completely
positive. We will refer to Eq. (\ref{causbound}) as the \emph{causality bound}
for a quantum operation $\map E_i$ in a generalized teleportation scheme.

We can now focus on the a probabilistic simulation of a given quantum channel
$\map C$ from the future to the past.  In this case we want that for one
particular measurement outcome $i_0 \in X$ the transformation $\map E_{i_0}$ is
proportional to $\map C$, namely $\map E_{i_0}  =  p  \map C$.  In this case,
the proportionality constant $p$ represents the probability that the simulation
succeeds.  The maximum value of the success probability $p$  must be compatible
with the causality bound
 \begin{equation}
             p\mathcal{C}\leq \rho_{0}~  \rm{Tr}_{in}. \label{MapsInequality}
 \end{equation}
 We will refer to the above bound as the \emph{causality bound for the
simulation of channel $\map C$}. 
 An obvious consequence of the causality bound is the following:  
 \begin{prop}
 A channel $\map C$ can be simulated with probability $p=1$ in a generalized
probabilistic teleportation protocol if and only if  $\map C$ is an erasure
channel, namely $\map C  = \rho_{0}  ~ \Tr_{in} $.    
 \end{prop}
 
 {\bf Proof.}  Suppose that we have $p=1$  in the causality bound of Eq.
(\ref{MapsInequality}) and define the quantum operation $\map D :  =  \rho_0  ~
\Tr_{in}  -  \map C$, so that we have $\map C  +  \map D  =  \rho_0  ~
\Tr_{in}$.  Applying both sides of the equality to a generic state $\rho$ and
taking the trace, we obtain  $\Tr[\map C(\rho)]  +  \Tr[\map D(\rho)] =
\Tr[\rho]$. Since $\map C$ is a quantum channel, by definition $\Tr[\map
C(\rho)]=\Tr[\rho]$. That is, $\Tr[\map D(\rho)]= 0$ for every $\rho$.  Hence,
we conclude that $\map D =
0$, and, therefore $\map C  =  \rho_0 ~ \Tr_{in}$.  $\blacksquare$ 
  
 In the following section we will show that the validity of the causality bound
is not only necessary, but also sufficient for the existence of a probabilistic
simulation with probability $p$. 
           
\subsection{Causality bound on Choi operators}
A very convenient tool to study quantum operations and channels is the Choi
correspondence,  which  associates  linear operators on he Hilbert space
$\mathcal{H}_{out}\otimes \mathcal{H}_{in}$  to linear maps sending operators on
$\mathcal{H}_{in}$ to operators on $\mathcal{H}_{out}$.  The Choi operator $E$
corresponding to a linear map $\mathcal{E}$ is given by 
\begin{equation}
E=\left(\mathcal{E}\otimes\mathcal{I}_{in}
\right)\left(|I\rangle\!\rangle\langle\!\langle
I|\right)\label{correspondence}\end{equation}
where $|I\rangle\!\rangle$ indicates the unnormalized maximally entangled
state $\sum_{i}|i\rangle\otimes|i\rangle$, $\{|i\>~|~  i=1, \dots, d_{in}\}$
being an orthonormal basis for $\hilb_{in}$. 

We will often use the ``double ket" notation of Ref.\cite{twoket-notation} for
bipartite
states on a tensor product Hilbert space $\mathcal{H}\otimes\mathcal{K}$:  This
notation is based on the one-to-one correspondence between operators $C$  from
$\mathcal K$ to $\mathcal H$ and states $|C\>\!\>$  in 
$\mathcal{H}\otimes\mathcal{K}$, given by 
\begin{align*}
|C\>\!\>  :  =  \sum_{i=1}^{d_{\mathcal H}} \sum_{j=1}^{d_{\mathcal K}}    ~  \<
 \varphi_i|  C  |\psi_j\>  ~  |\varphi_i\>  \otimes |\psi_j\>,
  \end{align*}  
where  $\{|\varphi_i\>\}_{i = 1}^{d_{\mathcal H}}$ is an orthonormal basis for
$\hilb$ and  $\{|\psi_j\>\}_{j = 1}^{d_{\mathcal K}}$ is an orthonormal basis
for   $\mathcal K$.  The ``double ket" notation has the useful properties  
\begin{align}\label{doubleketprod}
\<\!\<  C  |  D\>\!\>  =    \Tr[C^\dag D]~,
\end{align}
for  every pair of operators $C, D$ from $\mathcal K$ to $\mathcal H$, and 
 \begin{align}\label{doubleketalg}
 (A  \otimes B)  |C\>\!\>  =   |ACB^T  \>\!\>~,
\end{align}
for every operator $A$ on $\mathcal H$, $B$ on $\mathcal K$, and $C$ from
$\mathcal K$ to $\mathcal H$.  Here   $B^T: =  \sum_{i,j}   ~  \<\psi_i|  B 
|\psi_j\>  ~  |\psi_j\>\<\psi_i|$ denotes the transpose of $B$ in the given
basis   $\{|\psi_j\>\}_{j = 1}^{d_{\mathcal K}}$.

Since the correspondence between linear maps and Choi operators is one-to-one
every property of a linear map corresponds  to a property
of the associated Choi operator. 
In particular, we have Choi's theorem,  stating that a linear map $\map E$ is
completely
positive
if and only if its Choi operator $E$ is positive semi-definite \cite{choi}.
Moreoever, a linear map is trace preserving if and only if its Choi operator
obeys ${\rm Tr}_{out}[E]=I_{in}$. 

Let us now re-cast the  causality bound of Eq. (\ref{MapsInequality}) in terms
of Choi
operators.   The Choi
operator of the erasure channel $\map E_X  =  \rho_0  ~ \Tr_{in} $ is given by 
 \begin{align*}
\nonumber 
E_{X}&=  \left(\mathcal{E}_X \otimes\mathcal{I}_{in}
\right)\left(|I\rangle\!\rangle\langle\!\langle
I|\right)\\
&  =  \rho_0  \otimes I_{in}.  \label{eq:det-choi}
\end{align*}
Denoting by $C$  the Choi operator of the quantum channel $\mathcal{C}$,    the
causality  bound of  Eq. (\ref{MapsInequality})
becomes 
\begin{equation}
 pC\leq\rho_{out}\otimes I_{in} \label{ChoiInequality}.
\end{equation}
This operator form of the causality bound is much simpler to handle than the
version with linear maps.  In the following we will use Eq.  
(\ref{ChoiInequality}) to find the maximum probability $p$ for the simulation of
a channel $\map C$ to the past.  

We now use the method of quantum combs \cite{prlcombs,pracombs} to show that any
value of $p$ compatible with the causality bound of Eq.   (\ref{ChoiInequality})
    can be achieved in a suitable generalized teleportation protocol.
The key result that we use is Theorem 4 of Ref. \cite{pracombs}, whose
statement, specialized to the case  considered in this paper, can be
rephrased as
\begin{lemma}\label{lem:shortcomb}
Let $\{E_i\}_{i\in X}$ be a collection of positive operators on $\hilb_{out}
\otimes \hilb_{in}$ satisfying the property $\sum_{i\in X}  E_i  =   \rho_0 
\otimes I_{in}$ for some state $\rho_0$ on  $\hilb_{out}$.  
Then, there exists an ancillary system $A$, a pure state $\Psi \in  \hilb_A
\otimes \hilb_{out} $, and a POVM $\{F_i\}_{i  \in X}$ on $\hilb_{in} \otimes
\hilb_A$ such that, for every $i\in X$, $E_i$ is the Choi operator of the
quantum operation $\map E_i$ given by
\begin{align}
\map E_i   (\rho):  =  \Tr_{in, A}  [  (\rho \otimes |\Psi\>\<\Psi| )  (   F_i
\otimes I_{out})] ,  
\end{align} 
for every state $\rho$ on $\hilb_{in}$.
\end{lemma}

A simple consequence of lemma \ref{lem:shortcomb} is that every value of $p$
compatible with the causality bound is attainable in some generalized
teleportation protocol:  

\begin{theo}[Achievability of the causality bound]\label{theo:existence}  
Let $\map C\in  \Lin (\hilb_{out} \otimes \hilb_{in} ) $ be the Choi operator of
the  channel  $\map C$. If there exists a state $\rho_0$ on  $\hilb_{out}$ such
that $p C  \le  \rho_0 \otimes I_{in}$, then there exists a generalized
teleportation scheme such that  simulates $\map C$ with probability $p$.   
\end{theo}
{\bf Proof.}   To prove the thesis it is enough to define $E_0  :=  p C$  and
$E_1  :=  \rho_0  \otimes I_{in}  -  p C$ and to apply  lemma 
\ref{lem:shortcomb}  to the collection of operators $\{E_0, E_1\}$.  The
simulation of channel $\map C$ will succeed when the measurement outcome is $0$.
   \qed  

Thanks to theorem \ref{theo:existence} we can derive the maximum probability of
successful simulation from the causality bound of Eq. (\ref{ChoiInequality}): We
will know automatically that there exists a generalized teleportation protocol
succeeding with that maximum probability.    

\section{Ideal channels}\label{sec:ideal}
In this Section we will illustrate the general method of the causality bound in
two simple  examples.  The first example is about the probabilistic simulation
of an ideal classical channel, which preserves the elements of an orthonormal
basis, but destroys all off-diagonal elements of the density matrix.    In this
case, we will show that the probability of success is given by 
\begin{align*}
p_{cl}  =  \frac 1 d,
\end{align*}
where $d$ is the dimension of the Hilbert space.  
The second example is about the probabilistic simulation of an ideal quantum
channel, represented by the identity map.  In this case, using the  causality
bound   
 we will show that the probability of success is given by 
 \begin{align*}
p_q  =  \frac 1 {d^2},
 \end{align*}
 $d$ being again dimension of the Hilbert space.  This also proves that the
original teleportation protocol \cite{Original} is optimal for the probabilistic
simulation of an ideal channel from the future to the past.

\subsection{Optimal probabilistic simulation of an ideal classical channel from
the future to the past}\label{subsec:idealclassical}
\label{subsect:idealclassical}
An ideal classical channel can be modelled in Quantum Theory as a von Neumann
measurement on an orthonormal basis, represented by the channel $\map C_{cl} 
(\rho)  =    \sum_{i=1}^d   |i\>\<i  |  ~  \<  i|\rho |i\> $, where $\rho$ is
the state of a $d$-dimensional quantum system.  
  By the definition of Choi operator
 in Eq.  (\ref{correspondence}), the Choi operator of $\map C_{cl}$ is given by 
 \begin{equation}
 C_{cl}  =   \sum_{i=1}^d  |i\>\<i|  \otimes |i\>\<i| .   
 \end{equation} 
Inserting the above expression in the causality bound of Eq.
(\ref{ChoiInequality}), we obtain  
\begin{align*}
p  \left( \sum_{i=1}^d  |i\>\<i|  \otimes |i\>\<i|  \right) \le  \rho_{out } 
\otimes I_{in} .
\end{align*} Then,  taking on both sides the expectation value on the state
$|i\> \otimes |i\> $, we get
$p \le \< i| \rho_{out} |i\>   $, for every $i=1, \dots, d$,
 which also implies  $p \le \min_{1\le i \le d}  \<  i|\rho_{out}  |i\>  \le
\frac 1 d$.   Choosing $\rho_{out}  =  \frac {I_{out} } {d}$ we then achieve the
maximum value  
 \begin{align*}
 p_{cl}  = \frac 1 d.
 \end{align*}
A generalized teleportation protocol that simulates the ideal classical channel
with maximum probability $p_{cl}  =  \frac 1d $ can be easily obtained by
preparing the classically correlated state 
 \begin{align*}
 \sigma_{cl}  = \frac {\sum_i   |i\>\<i|  \otimes |i\>\<i| }d
 \end{align*} 
 and by measuring the two-outcome POVM $\{P_0, P_1\}$ given by $P_0  := \sum_{i}
  |i\>\<i|  \otimes |i\>\<i| $ and $P_1  :=  I  -  P_0$.

\subsection{Optimal probabilistic simulation of an ideal quantum channel from
the future to the past}\label{sec:Original}
Another instructive application of the general method is the  proof that  the
value of the probability of successful teleportation $p_q=   \frac{1}{d^2}$ in
the
original
teleportation protocol \cite{Original}  is the optimal value determined by
causality.

In the case of teleportation,   the quantum channel that we want to simulate is
just 
the identity: $\mathcal{C}_q=\mathcal{I}$. By definition of Choi operator
 Eq.  (\ref{correspondence}), the Choi operator of the identity channel is
\begin{equation*}
C_q= |I\rangle\!\rangle\langle\!\langle I|. \end{equation*}
Plugging this into the causality bound of Eq.  (\ref{ChoiInequality}), we then
obtain
 \begin{equation}                                        
p  |I\rangle\!\rangle\langle\!\langle I|\leq \rho_{out}\otimes
I_{in}.\label{OriginalIneq}
                                                      \end{equation}
To find the maximum value of the simulation probability $p$ we must ensure that
the difference between the right hand side and the left hand side  is a positive
operator.    In general, this is not a trivial issue because the operators on
the two sides are not necessarily
diagonalizable in the same basis. 

To resolve this problem, we notice a 
symmetry of the left hand side. Defining the complex conjugate of an operator as
$A^*  : =  \sum_{i,j}   \< i|  A  |j\>^*  ~ |i\>\<  j|$ and using Eq.
(\ref{doubleketalg}) we have 
 \begin{align*}
  (U\otimes U^{*})  |I\rangle\!\rangle=|I\rangle\!\rangle  \qquad \forall U \in
SU(d).
  \end{align*} 
 and, therefore, $(U\otimes U^{*})
  C_q (U\otimes U^{*})^\dag  =  C_q  \quad \forall U \in SU(d)$. 
  Applying the same transformation on
both sides of the inequality (\ref{OriginalIneq}), we obtain
\begin{align*}
p|I\rangle\!\rangle\langle\!\langle I|\leq U\rho_{out} U^{\dagger}\otimes I_{in}
 \qquad \forall U \in SU(d).
\end{align*} 
Since the inequality must hold for every unitary $U\in SU(d)$, it also holds for
the
integral over the normalized Haar measure $d U$, \begin{align*}
p|I\rangle\!\rangle\langle\!\langle I|  &\leq
\int d U ~  \left(  U\rho U^{\dagger}\otimes I_{in}  \right)  \\
&  =  \frac {I_{out} \otimes I_{in}} d,
                                                               \end{align*}
                                                               having used the
Schur's lemma for the last equality.
                                                               Now, both sides
of the inequality are diagonalizable in the same basis. Since the operator on
the left hand side  has eigenvalue $ p d$ and the operator on the right hand
side
has eigenvalue
of $\frac{1}{d}$,  the inequality becomes $p d\leq \frac{1}{d}$. Hence, the
maximum
probability of teleportation compatible with the causality bound is
\begin{equation*}
                             p_{q}=\frac{1}{d^2}
                            \end{equation*}
Thus, we have shown from causality that the original teleportation
protocol is the optimal probabilistic simulation of an identity channel from the
future to the past.  

\subsection{Lower bound for general channels} 

The probability of success for ideal channels provides a lower bound for the
probability of success for arbitrary channels: Indeed, if we have protocols
that simulate  ideal channels from the future to the past, then we can use these
protocols to simulate a desired channel $\map C$ in two ways:  \emph{i)} we can
apply  $\map C$ to the input in the future and then send the output to the past
through the ideal channel, or \emph{ii)}  we can send the input to the past
through the ideal channel and apply the desired channel $\map C$ to the input in
the past.  This argument leads to the following:  

\begin{prop}
Let $\map C$ be a quantum channel sending states on the Hilbert space
$\hilb_{in}$ to states on the Hilbert space $\hilb_{out}$.  Then, the maximum
probability $p_{\map C}$  to simulate the action of  $\map C$ from the future to
the past satisfies the bound
\begin{align}\label{generallower}
p_{\map C}  \ge  \max\left \{  \frac 1{d^2_{in}} , \frac 1{d^2_{out}}  \right\}.
\end{align}
For classical channels, of the form $\map C(\rho)  =  \sum_{i,j}     p (j|i)  ~
|\psi_j\>\<\psi_j|     \< \varphi_i|  \rho  |\varphi_i\>$ where  
$\{|\varphi_i\>\}_{i=1}^{d_{in}}$ and  $\{|\psi_j\>\}_{j=1}^{d_{out}}$ are
orthonormal bases for $\hilb_{in}$ and $\hilb_{out}$, respectively, the bound
becomes
\begin{align}\label{generallower}
p_{\map C}  \ge  \max\left \{  \frac 1{d_{in}} , \frac 1{d_{out}}  \right\}.
\end{align}
\end{prop}

\section{Measure-and-prepare channels}\label{sec:measprep}

In this section we focus on measure-and-prepare channels, namely channels $\map
C_{m\&  p}$ of the form  
\begin{align}\label{eq:mp}
\map C_{m\&  p }  (\rho)  =  \sum_{i \in  X}         \rho_i   ~  \Tr [ P_i  
\rho   ] ~,
\end{align}
where $\{P_i\}_{i\in  X}$   is a POVM on the input Hilbert space $\hilb_{in}$,
that is, 
\begin{align*}
P_i\ge 0   \quad \forall i \in X,   \qquad \sum_{i \in  X}  P_i  =  I_{in}
\end{align*}   
and $\{\rho_i\}_{i \in X}$ is a collection of normalized density matrices on the
output Hilbert space $\hilb_{out}$, that is,    
\begin{align*}
\rho_i\ge 0,   \qquad \Tr[ \rho_i]  =  1 \quad \forall i \in  X~.
\end{align*}   
Measure-and-prepare channels are interesting as an example of channels that
transmit only classical information: Indeed, it is well known that the quantum
capacity of a measure-and-prepare channel is always zero (this follows, e.g.
from the general upper bound presented by Holevo and Werner in Ref. 
\cite{holwer}).   

In this section we will first show that the maximum probability to simulate a
given 
  measure-and-prepare channel from the future to the past satisfies the general
lower bound  
\begin{align}\label{mpbound}
p_{m \&  p} \ge\frac{1}{d_{out}},
\end{align}  
where $d_{out}$ is the dimension of the output Hilbert space. 
The bound is achieved for $\hilb_{in}  =  \hilb_{out}$ by von Neumann
measurements on an orthonormal basis, and in that case coincides with the bound
for the simulation of an ideal classical channel (cf. subsection 
\ref{subsect:idealclassical}).  Eq. (\ref{mpbound})  is quite intuitive:  If a
channel can only transmit classical information, then the probability to
simulate  it is at least equal to the probability  to simulate an ideal
classical channel with the same output.

Then we will compute the exact value of the probability of success for two 
particular examples of measure-and-prepare channels:  
The first example will be the channel associated to the optimal estimation of a
pure state, namely the linear map $\map C_{est}$ sending states on $\hilb$ to
states on $\hilb$ given by
\begin{align}\label{Cest}
\nonumber \map C_{est}  (\rho)  =    \int d  U  ~      \rho_U    ~  \Tr[  P_U 
\rho  ]  \qquad  &\rho_U:  =  U |0\>\<0| U^\dag\\
  &       P_U : =   d ~ U |0\>\<0| U^\dag~,
\end{align} 
where $d$ is the dimension of $\hilb$ and $|0\>  \in \hilb$ is a fixed state.  
In this case the general bound is tight:  the probability of success is given
by 
\begin{align*}
p_{est}  =  \frac 1 d. 
\end{align*}
The second example will consist in the \emph{universal NOT} channel \cite{UNOT},
namely the
linear map   $\map C_{U-NOT}$ sending states on $\hilb$ to states on $\hilb$
given by  
\begin{align}\label{unot}
\nonumber \map C_{U-NOT}  (\rho)   =    \int  d U  ~       \rho^{\perp}_U       
~  \Tr[  P_U  \rho] \qquad   &\rho^{\perp}_U   :=    \frac{U (I  -  |0\>\<0|) 
U^\dag}{d-1}   ~~ \\    
&P_U  := d ~   U  |0\>\<0  |  U^\dag         ~~    
\end{align} 
In this case we will show  that the probability of success is given by 
\begin{align*}
p_{U-NOT}  =  1-  \frac 1 {d^2},
\end{align*} 
well above the lower bound $p_{m\& p}  \ge  \frac 1 d$.    The intuition that
the probability of success depends on the amount of information that a channel
can transmit suggests that the universal NOT channel can  transmit less
classical information than the state estimation channel.

 \subsection{General lower bound on the probability of success}
 The Choi operator of the generic measure-and-prepare  channel  $\map C_{m \&
p}$  in Eq. (\ref{eq:mp}) is given by
  \begin{align*}
C_{m\&p} & =\sum_{i\in X}  \left(  \rho_{i}\otimes P_{i}^{T} \right).
\end{align*}
Let us consider the causality bound $p~ C_{m\& p}  \leq\rho_0 \otimes I_{in}$,
where $\rho_0$ is a state on $\hilb_{out}$ to be optimized.   
A (possibly suboptimal)  choice for the state $\rho_0$  is 
$\rho_0=\frac{I_{out}}{d_{out}}$,  for which the bound becomes  
\begin{align*}
p~ C_{m\& p}  \leq\frac{I_{out} \otimes I_{in}}{d_{out}}.
\end{align*} 
The bound will be satisfied as long as we choose $p   =   \frac1  {  d_{out}  
\gamma_{\max}}  $, where $\gamma_{max}$ is the largest eigenvalue of the Choi
operator $C_{m\& p}$.  

We now show that $\gamma_{max} $ cannot be larger than $1$.  To this purpose, we
consider  an arbitrary normalized state 
$|\Psi \Rangle \in \hilb_{out}\otimes\hilb_{in}$ and show that $\Langle \Psi  | 
C_{m\&  p}  |\Psi \Rangle  \le 1$.   
Using Eq. (\ref{doubleketprod}), the normalization of the state $|\Psi\>\!\>$ 
reads 
\begin{align}\label{psinorm}
\Langle\Psi|\Psi\Rangle = \rm{Tr}[\Psi^{\dagger}\Psi]=1.  
\end{align}
 We then obtain
\begin{align*}
                                           \Langle\Psi|     C_{m \& p}   
|\Psi\Rangle &= \sum_{i\in X} \Langle\Psi|\rho_i \Psi P_i\Rangle\\
&=\sum_{i\in X} \Tr[\Psi^{\dagger}\rho_i \Psi P_i].\\
&=\sum'_i    \Tr[ \Psi P_i  \Psi^{\dagger} ]   \Tr[  \rho_i  \sigma_i ] \quad 
\sigma_i: =  \frac{ \Psi P_i \Psi^{\dagger} }{\Tr[\Psi P_i \Psi^{\dagger}]}\\ 
& \le \sum_{i \in X} \Tr[ \Psi P_i \Psi^{\dagger} ] \\
&  =  \Tr[\Psi^\dag \Psi]\\
& =1.
\end{align*}
Here we used Eq. (\ref{doubleketalg}) for the first equality and  Eq.
(\ref{doubleketprod})  for the second.  The sum in the third equation is
restricted to the values of $i$ such that $\Tr[ \Psi P_i \Psi^{\dagger} ]   \not
 = 0$. The inequality is due to the fact that the overlap between the density
matrices $\rho_i$ and $\sigma_i$ is bounded by $1$,  the fourth equality is due
to the normalization condition $\sum_{i \in X} P_i  = I_{in}$, and the last
equality is just Eq. (\ref{psinorm}).    

Since the bound $\Langle \Psi  |  C_{m\&  p}  |\Psi \Rangle  \le 1$ holds for
arbitrary $|\Psi\Rangle$, we conclude that the maximum eigenvalue of $C_{m \&
p}$ satisfies $\gamma_{\max} \le 1$. This implies that the value $p=    1/
{d_{out}}$ is compatible with the causality bound, and therefore, it is
achievable with a suitable generalized teleportation protocol.       Hence, for
the maximum probability $p_{m\&  p}$ we have the lower bound  
$p_{m\& p}  \ge \frac 1 {d_{out}}$.

\subsection{Optimal pure state estimation}
The Choi operator for the pure state estimation channel is 
\begin{align*}
C_{est }  &  = d \int d U ~ ( U \otimes U^*)   (|0\>\<0|  \otimes |0\>\<0|)    
( U^\dag \otimes U^T)\\
   &  =  \frac 1 d  |I\>\!\>\<\!\<  I|   +     \frac{1}{d+1}   \left(  I 
\otimes I  -   \frac {|I\>\!\>\<\!\<  I| } d   \right).       
\end{align*}
Clearly, $C_{est}$ enjoys the symmetry  $ ( U \otimes U^*)   C_{est}   ( U^\dag
\otimes U^T)  =   C_{est} $ for every $U \in SU (d)$, which implies that the
causality bound has the form 
\begin{align*}
p ~  C_{est}  \le \frac {I \otimes I}{d}
\end{align*}
(the proof is the same as the proof for the ideal quantum channel in subsection 
\ref{sec:Original}).  

Since the maximum eigenvalue of $C_{est}$  is $\gamma_{\max}  = 1$, we have that
the maximum probability is $p_{est}=  \frac 1 d$.  

\subsection{Universal NOT channel}
Using Eqs. (\ref{Cest})  and (\ref{unot}),  the universal NOT channel can be
written as 
\begin{align*}
\map C_{U-NOT}    =  \frac 1 {d-1}   \left (     I ~  \Tr     -   \map C_{est} 
\right).             
\end{align*}
Hence, its Choi operator is given by 
\begin{align*}
C_{U-NOT}  & =        \frac 1 {d-1}   \left (     I  \otimes I     -   C_{est} 
\right)\\
  &  =  \frac d {d^2  -1}    \left(  I \otimes I  -    \frac{|I\>\!\>\<\!\<I|} d
  \right).  
\end{align*}
Again, we observe that $C_{U-NOT}$ enjoys the symmetry  $ ( U \otimes U^*)  
C_{U-NOT}   ( U^\dag \otimes U^T)  =   C_{U-NOT} $ for every $U \in SU (d)$,
which implies that the causality bound has the form 
\begin{align*}
p ~  C_{U-NOT}  \le \frac {I \otimes I}{d}.
\end{align*}
Now, the maximum eigenvalue of $C_{U-NOT}$ is  $\gamma_{\max}  = \frac d {d^2 
-1} $. Hence,  the maximum probability is given by $p_{U-NOT}  =  \frac 1 {d 
\gamma_{\max}}  =  1 -  \frac 1 {d^2}$.

\section{Optimal probabilistic teleportation with multiple input and output
copies}\label{sec:opttele}
In this section we consider probabilistic protocols that take $N$ input copies
of a pure state in the future and produce $M\le N$ copies of the same state in
the past.  
In this case, the channel that we want to simulate is the partial trace over
$(N-M)$ systems, symmetrized over all possible choices of the systems that are
traced:  
\begin{align}\label{symmtrace}
\map C_{N \to M}   (\rho) =     \Tr_{N}  \Tr_{N-1}  \dots \Tr_{M+1}     \left[ 
\frac 1 {N!}   \sum_{\pi  \in  S_N}     U^{(N)}_\pi \rho  U^{(N) 
\dag}_\pi\right],    
\end{align}
Here the partial trace is over all input Hilbert spaces except the first $M$,
$\pi$ is an element of the group $S_N$ of the permutations of $N$ objects and
$U^{(N)}_\pi$ is the unitary operator that permutes the $N$ input Hilbert spaces
according to the permutation $\pi$.

 The goal will be to see how the probability of success, denoted by $p^+_{q, N
\to M}$, varies with the number of input and output copies. In particular, we
will show that the probability of success increases with the number of available
input copies.   
 
 Note that in the classical world there cannot be any increase of the
probability of success with the number of copies of an pure input state:  Since
we can perfectly
clone the classical pure states corresponding to an orthonormal basis, we must
have  
\begin{align}
p^+_{cl, N  \to M}  =     p^+_{cl, 1 \to  1}    =  \frac 1 d \qquad \forall M,
N.   
\end{align} 

The classical value $p_{cl}  =  \frac 1 d$ is clearly an upper bound for the
probability of success in the quantum case:  Indeed, if we have a probabilistic
teleportation protocol that  succeeds with probability $p^+_{q, N \to M}$  in
transforming $N$ copies of a pure quantum state into $M$ copies of the same
quantum state, then the protocol will work in particular on the states of an
orthonormal basis,  thus simulating  the classical trace channel from $N$ to $M$
copies.      Hence, we must have 
\begin{align}
p^+_{q, N \to M}   \le  \frac 1 d \qquad \forall N, M .
\end{align}  
This bound is very intuitive:  If there existed a quantum protocol that
succeeded with probability $p^+_{q,  N \to M}  > \frac 1 d$, then two parties,
Alice and Bob, could use the protocol to reliably win a lottery where the goal
is to guess today a winning number $n$ between 1 and $d$ that will be extracted
tomorrow.   To win the lottery, Alice and Bob can agree to use an orthonormal
basis $\{|n\>\}_{n=1}^d$ such that
the basis vector $|n\>$  corresponds to the number $n$.  Tomorrow, when the
winning number is announced, Bob inputs  into the teleportation machine $N$
identical systems,  each of which is prepared in the state $|n\>$ corresponding
to  the winning number.
Today Alice measures one of the $M$ copies in the agreed basis and with
probability $p^+_{q, N \to M}$ she obtains the winning number, just in time to
submit  her lottery ticket. If  $p_{q, N \to M}^+$ were larger than the
classical value $p_{cl} =  1/d$, then Alice would
win the lottery with probability larger than the probability $p_{ran}  =  \frac
1 d$ to win the lottery by a random guess. 


In the following we will show that, unlike in the classical case, in the quantum
case the probability of successful teleportation increases with the number of
input copies.  
First, we will consider the simplest case $N=2, M=1$, where the calculation of
the maximum probability of success can be done explicitly even without
constraining the input copies to be in a pure state.  In this case, we find out
that the  probability increases from the value  $p_{1 \to 1}  = \frac{1}{d^2}$ 
in the $N=1$ case to the value $p_{2 \to 1} = \frac 2 {d(d+1)}$.    This result
suggests that the quantum improvement is a general feature that does not depend
on the assumption of pure input states.
 We will then move to the case of general $N$, keeping the number of output
copies fixed to $M=1$.  In this case, we will that
the success probability approaches the classical value  $p_{cl}  = \frac{1}{d}$
as $N \to\infty$.  An explanation of this asymptotic behaviour  will be provided
in Sec. \ref{sec:summary}.


Finally, we will extend the analysis to general $M$, also including the case
$M>N$, where optimal clones of the input state are
teleported to the past.  Surprisingly, in the case of cloning channels we will
show that for fixed $N$  the probability of success is an increasing function of
the number of output copies  $M$.  Also this fact will be explained in Sec.
\ref{sec:summary}.

\subsection{From two copies of a mixed state in the future to one copy in the
past}
For two copies of input and one copy of output, the input Hilbert space is
$\mathcal{H}_{in}=\mathcal{H}^{\otimes 2}$ and the output Hilbert space is
$\hilb_{out}  = \mathcal{H} \simeq \mathbb C^d$.  In this case, the quantum
channel to be simulated is \begin{align*}
\mathcal{C}_{q, 2\to
1}=\frac{1}{2}\left(\mathcal{I}_{1}\rm{Tr}_{2}+\mathcal{I}_{2}\rm{Tr}_{1}
\right),
\end{align*}
where the labels  1 and 2 label on the right hand side label the two input
spaces.  Using Eq.  (\ref{correspondence}), we find that the Choi operator of
this channel is given by 
\begin{align*}
C_{q, 2\to
1}
&=\frac{1}{2}\left(|I\Rangle\Langle I|_{01}\otimes I_{2} + |I\Rangle\Langle
I|_{02}\otimes I_{1}\right),                                          
\end{align*}
 where we have used 0 to label the output Hilbert space. Plugging this into
the causality bound of Eq. (\ref{ChoiInequality}) we obtain
\begin{equation}\label{oramuoio}                
p\frac{1}{2}\left(|I\Rangle\Langle I|_{01}\otimes I_{2} + |I\Rangle\Langle
I|_{02}\otimes I_{1}\right)\leq \rho_{0}\otimes I_{12} 
                                              \end{equation}

Like in subsection \ref{sec:Original}, we can notice a symmetry of the Choi
operator: 
\begin{align*}
(U\otimes U^{*}\otimes U^{*})  C_{q, 2\to
1}   ( U^{\dagger}\otimes U^{T}\otimes U^{T}) = C_{q, 2\to1} ~,
\end{align*} 
for every  $U \in SU(d)$.
Applying this transformation to both sides of the inequality  (\ref{oramuoio})
and integrating over the Haar measure  we obtain
\begin{equation*}                
p\frac{1}{2}\left(|I\Rangle\Langle I|_{01}\otimes I_{2} + |I\Rangle\Langle
I|_{02}\otimes I_{1}\right)\leq \frac{1}{d} I_{0}\otimes I_{12}.
\label{2to1Ineq}
                                              \end{equation*}
It now remains to find the eigenvalues of the Choi operator $C_{q, 2\to1}$. To
do
this, we define the following set of orthonormal vectors in
$\hilb_{out}\otimes\hilb_{in}$: \begin{align*}               
|n_{+}\rangle
&=\frac{1}{\sqrt{2(d+1)}}\left(|I\Rangle_{01}|n\rangle_{2}+|I\Rangle_{02}
|n\rangle_ { 1 } \right) \\
|n_{-}\rangle
&=\frac{1}{\sqrt{2(d-1)}}\left(|I\Rangle_{01}|n\rangle_{2}-|I\Rangle_{02}
|n\rangle_{1} \right). 
                  \end{align*}
In terms of these vectors, the Choi operator can be rewritten  as
\begin{equation}
 C_{q, 2 \to1}=\frac{d+1}{2}\sum_{n}|n_{+}\rangle\langle
n_{+}|+\frac{d-1}{2}\sum_{n}|n_{-}\rangle\langle
n_{-}|
\end{equation}
Thus we see that the eigenvalues of this Choi operator are $\gamma_{max} =
\frac{d+1}{2}$ and
$\gamma_{\min} = \frac{d-1}{2}$. To ensure that Eq.  (\ref{oramuoio}) holds, we
must have  $p \gamma_{max} \leq\frac{1}{d}.$ Thus, the maximum
teleportation probability from 2 copies of input is 
\begin{equation*}          
p_{q, 2\to1}=\frac{2}{d(d+1)},
\end{equation*}
which is greater than the maximum probability of teleportation from 1 to 1 copy,
$p_{q, 1 \to 1}  =\frac{1}{d^2}$. In other words,  adding an extra copy of the
input yields
improvement in the teleportation probability.  

It is worth stressing that, since we are simulating the symmetrized trace
channel $\map C_{q, 2\to 1}$, any bipartite state $\sigma$ on  $\hilb_{in}$
with the property $   (\Tr_{1}[\sigma]  +  \Tr_{2} [\sigma] )/2  = \rho $ will
be transformed with probability $  p_{q, 2\to1}=\frac{2}{d(d+1)}$ into $\rho$.  
This holds in particular when $\sigma =  \rho \otimes \rho$, corresponding to
two identical copies of the  mixed state $\rho$.   

\subsection{From $N$ copies of a pure state in the future to $M= 1$ copy in the
past}
We now calculate the maximum probability of teleportation from $N$
input copies to one output copy. In this case, we will only require the
teleportation protocol to work perfectly when we are inputting $N$
 copies of the same pure state. 
Since the density matrix of $N$ identical copies of a pure state
has support in the symmetric subspace of $\mathcal{H}^{\otimes N}$, henceforth
denoted
as $\left(\mathcal{H}^{\otimes N}\right)_{+}$, we will focus on the restriction
of the trace channel to the
symmetric subspace, namely on the channel $\map C_{q, N\to 1}^+$ given by 
\begin{align}
\map C_{q, N\to 1}^+   (\rho) :  =  \map C_{q, N \to 1}   \left(  P_+^{(N)}    
\rho  P_{+}^{(N)}  \right) ,  
\end{align}
where $P^{(N)}_+$ is the projector on $\left(\mathcal{H}^{\otimes N}\right)_{+}$
and $\map C_{q,N\to 1}$ is the symmetrized partial trace channel defined in Eq.
(\ref{symmtrace}). Note that the map $\map C_{q, N\to 1}^+$ is trace-preserving
only for states with support in $\left(\mathcal{H}^{\otimes N}\right)_{+}$.

It is useful to observe that, due to the projection on the symmetric subspace, 
the sum over all permutations in Eq. (\ref{symmtrace})  can be omitted:  Indeed,
we have
\begin{align*}
&\map C_{q, N\to 1}^+   (\rho) = \\
&\qquad    =            \Tr_{N}  \dots \Tr_{2}     \left[  \frac 1 {N!}  
\sum_{\pi  \in  S_N}     U^{(N)}_\pi        P_+^{(N)}     \rho  P_{+}^{(N)}    
U^{(N)  \dag}_\pi\right]  \\
&\qquad    =            \Tr_{N}   \dots \Tr_{2}      \left[       P_+^{(N)}    
\rho  P_{+}^{(N)}   \right],        
\end{align*}
having used the property $   U^{(N)}_\pi        P_+^{(N)}     =       P_+^{(N)} 
   U^{(N)  \dag}_\pi  =      P_+^{(N)}    $  for every permutation $\pi\in S_N$.

\begin{figure}[h]
\begin{centering}
\includegraphics[scale=0.3]{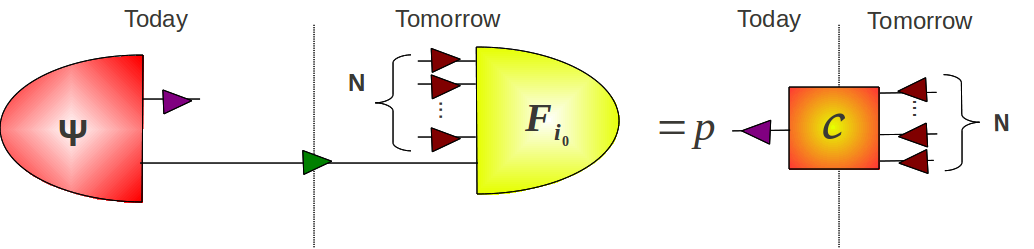}
\par\end{centering}

\caption{\label{fig:ncopies} (Color online) Probabilistic teleportation from $N$
copies to
$M=1$ copy of  a pure state. Here, $\mathcal{C}$ represents the symmetrized
partial trace over all but one
input copy, restricted to the symmetric subspace.}
\end{figure}

 

By the definition of Choi operator in Eq. (\ref{correspondence}) we find that
the Choi operator $C^+_{q, N \to 1}$ is given by  
\begin{align*} C^+_{q,N \to1} & =     ( I_{0}  \otimes P^{(N)}_+  )    \left(  |
I\rangle\!\rangle\langle\!\langle I|_{0  1}\otimes P_+^{ (N-1)}  \right)   (
I_{0}  \otimes P^{(N)}_+  ) ,
\end{align*}
where  the output Hilbert space is labelled by $0$, the input Hilbert spaces are
labelled with numbers from $1$ to $N$, and the projector $P_+^{ (N-1)} $ acts on
the tensor product  of all input  Hilbert spaces except the first.  


We want to find the maximum value  $p_{q, N \to 1}$ compatible with the
causality bound of Eq. (\ref{ChoiInequality}), which now reads
\begin{align}\label{eq:causalN1}
p ~    C_{q, N \to 1}^+  \le \rho_{out}  \otimes I_{in}.   
\end{align}

To this purpose, we notice that the Choi operator $C^+_{q, N \to 1}$ has the
symmetry 
\begin{align*}
(U\otimes U^{*\otimes
N}) C^+_{q, N \to 1}(U^{\dagger}\otimes U^{T\otimes N} )=C^+_{q ,N \to 1}~, 
\end{align*}
for every $ U \in SU(d)$.
Following again the group theoretic
argument of subsection \ref{sec:Original} we can rewrite Eq. (\ref{eq:causalN1})
as 
\begin{equation}
p   ~  C^+_{q, N\to 1}\leq\frac{1}{d}I_{out}\otimes
I_{in}.\label{eq:ineq-Nto1}\end{equation}
Hence, it only remains to find the eigenvalues of $C^+_{q, N \to 1}$.


Expanding the projector $P^{(N-1)}_{+}$ as 
\begin{align*}
P^{(N-1)}_{+}  =  \sum_{n=1}^{d_+^{(N-1)}}    |\varphi_n  \>  \< \varphi_n |~,
\end{align*} 
where $\{|\varphi_n\>\}_{n=1}^{d_+^{(N-1)}}$ is an orthonormal basis for $\left(
\hilb^{\otimes (N-1)}\right)_+$,   we can
express $C_{q, N\to 1}^+$ as
\begin{align}
\nonumber & C^{+}_{q,N\to1}= \\
\nonumber & \quad   =  ( I_0  \otimes P^{(N)}_+  )   \left(
\sum_{n=1}^{d_+^{(N-1)}}   |I\rangle\!\rangle \langle\!\langle
I|_{01}  \otimes 
|\varphi_n \rangle   \langle\varphi_n |  \right)     ( I_0  \otimes P^{(N)}_+  )
 \\
\label{Nto1diag}& \quad   =  \sum_{n=1}^{d_+^{(N-1)}}    |\Phi_n  \>\<  \Phi_n 
|    , 
\end{align}
having defined
\begin{align*}
  |\Phi_{n}  \>  :  = ( I_0  \otimes P^{(N)}_+  )    |I\rangle\!\rangle_{01}
|\varphi_n \rangle .  
\end{align*}
Eq. (\ref{Nto1diag})  is the desired diagonalization of the Choi operator
$C^+_{q ,N \to 1}$: Indeed, we have 
\begin{align*}
\<\Phi_{n}  | \Phi_{m}\>  & 
 =   \< \!\<  I|_{01}  \<  \varphi_n|   \left (I_0  \otimes P^{(N)}_{+}  
\right)    |I\>\!\>_{01} |\varphi_m\>  \\
  &  =   \<  \varphi_n|      \Tr_1 \left [P^{(N)}_{+}  \right]     |\varphi_m\> 
\\
  &  = \frac{d^{(N)}_+}{d^{(N-1)}_+}   ~     \<  \varphi_n|       P^{(N-1)}_{+} 
   |\varphi_m\>\\
   & =  \frac{d^{(N)}_+}{d^{(N-1)}_+}  ~ \delta_{n, m}  , 
\end{align*}
that is, the vectors $\{|\Phi_n\> \}_{n=1}^{d_+^{(N-1)}}$ are mutually
orthogonal and  have the same norm.  Therefore, $C^+_{q, N \to 1}$ only has
one non-zero eigenvalue, given by  $\gamma_{N\to 1} =
\frac{d^{(N)}_+}{d^{(N-1)}_+}$.     
Plugging this value into the causality bound of Eq.  \eqref{eq:ineq-Nto1} we
obtain $p  ~  \gamma_{N \to 1}  \le  \frac 1 d$. The maximum probability of
success is then given by
\begin{align*}
p_{q, N\rightarrow1} &=    \frac{d^{(N-1)}_+}{ d  ~d^{(N)}_+   }\\
  &  =    \frac{N}{d(d-1+N)}.
\end{align*}
Note that the probability of successful teleportation increases with the number
$N$ of input copies, unlike in the classical case. 
For $N=1$ we retrieve the value $p_{q}  =  \frac 1 {d^2}$ of the standard
teleportation protocol, while  in the limit of $N$ going to infinity  we observe
that $p_{q, N \to 1}$ tends towards the classical limit $p_{cl}  =\frac{1}{d}$.
Such an asymptotic behaviour will be explained in section \ref{sec:summary}.

\subsection{From $N$ copies of a pure state in the future to $M\le N$ copies of
the same state in the past}\label{subsect:traceNM}
 Here we calculate the maximum probability of teleportation from $N$ input
copies to $M\le N$ output copies of a generic pure state. Again, since for every
integer number $k$ the density matrix of $k$  identical copies
of a pure state has support in the symmetric subspace $\left(  \hilb^{\otimes k}
\right)_+$,  we will restrict the input of the partial trace channel $\map C_{q,
N \to  M }$ to be in $\hilb_{in} := \left(  \hilb^{\otimes N} \right)_+$  and
the output to be in $\hilb_{out}  : =\left(  \hilb^{\otimes M} \right)_+$.   
 In other words, we will focus on the probabilistic simulation of the channel
$\map C_{q, N \to M}^+$  given by 
 \begin{align*}
 \map C_{q, N \to M}^+  (\rho) :  =     P^{(M)}_+     \left[ \map C_{q, N \to M}
 \left (  P^{(N)}_+    \rho   P^{(N)}_+      \right) \right]    P^{(M)}_+ ~,
 \end{align*}
 where $\map C_{q,N\to M}$ is the symmetrized partial trace channel defined in
Eq.
(\ref{symmtrace}).
 
Again, thanks to the projection on the symmetric subspace we can write
\begin{align}\label{CNM}
 \map C_{q, N \to M}^+  (\rho) :  =     P^{(M)}_+     \left( \Tr_N  \dots 
\Tr_{M+1}  \left [  P^{(N)}_+    \rho   P^{(N)}_+      \right] \right)   
P^{(M)}_+ ~.
 \end{align}

\begin{figure}[h]
\begin{centering}
\includegraphics[scale=0.3]{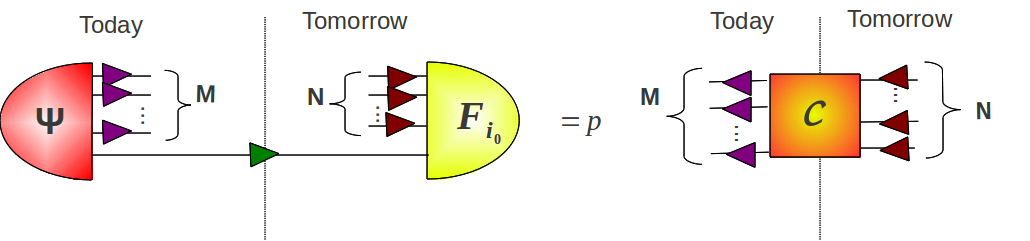}
\par\end{centering}

\caption{\label{fig:ntom} (Color online) Probabilistic teleportation from with
$N$ input copies
in the future to  $M\le N$ output
copies in the past. Here, $\mathcal{C}$ represents the symmetrized partial trace
over all but
$M$ of the $N$ input systems, where the input and output are constrained to be
in the symmetric subspace.}
\end{figure}



By the definition of Choi operator  in Eq. (\ref{correspondence}),  we then
obtain 
 \begin{align}\label{CNMlast}
 C^{+}_{q,N \to M}&=    \left(P_{+}^{(M)} \otimes  P^{
(N)}_+ \right)     K      \left(P_{+}^{(M)} \otimes  P^{(N)}_+ \right)   \\ 
 \nonumber K&=   \left(\prod_{i=1}^{M}
|I\rangle\!\rangle\langle\!\langle I|_{i i}\otimes
P_{+}^{(N-M)}\right) ,
\end{align}
where the projector $P_{+ }^{(N-M)}$  acts on the tensor product of the last
$N-M$ input Hilbert spaces.  


Our goal is to find the maximum probability of success compatible with the
causality bound
\begin{align*}
p  ~  C_{q, N \to M}^+   \le   \rho_{out}  \otimes I_{in},
\end{align*}
where $\rho_{out}$ is a suitable state on $\hilb_{out}  =\left(  \hilb^{\otimes
M} \right)_+$ and $I_{in}=  P^{(N)}_+$ is the identity on $\hilb_{in}   =\left( 
\hilb^{\otimes N} \right)_+$.  
Now, it is easy to see that the Choi operator $C^{+}_{q, N-M}$ has the symmetry
\[\left(U^{\otimes M}\otimes
U^{*\otimes N}\right)C^{+}_{q,N-M}\left(U^{\dagger\otimes M}\otimes U^{T\otimes
N}\right)=C^{+}_{q,N-M},\]
for every $U \in SU(d)$.   Due to this symmetry, the state $\rho_{out}$ in the
causality bound can be chosen without loss of generality to be the
invariant state  $\rho_{out}  =  \frac{I_{out}}{d_{out}}  = 
\frac{P^{(M)}_+}{d^{(M)}_+}$, so that the bound becomes 
 \begin{align}\label{boundNM}
p   ~  C_{q, N \to M}^+   \le     \frac{  P^{(M)}_+  \otimes P^{(N)}_+}
{d^{(M)}_+}  .
\end{align}

To find the maximum probability it is enough to find the maximum eigenvalue of
$C_{q, N \to M}$.  
Expanding the projector $P^{(N-M)}_{+}$ as  
\begin{align*}
P^{(N-M)}_{+}  =  \sum_{n=1}^{d_+^{(N-M)}}    |\varphi_n\>\< \varphi_n|
\end{align*} 
for some orthonormal basis $\{    |\varphi_n\>\}_{n  =1}^{d_+^{(N-M)}}$ for
$\left( \hilb^{\otimes (N-M)} \right)_+$,  we can
rewrite Eq. (\ref{CNMlast}) as    
\begin{align}
\nonumber 
C^{+}_{q,N \to M}&= \sum_{n=1}^{d_+^{(N-M)}}     P_{+}^{M\otimes
N} \left(     \bigotimes_{i=1}^{M} |I\rangle\!\rangle\langle\!\langle I|_{i
i}\otimes  |\varphi_n\>\< \varphi_n|      \right)P_{+}^{M\otimes N}\\
&= \sum_{n=1}^{d_+^{(N-M)}}      |\Phi_n\>\<\Phi_n|~,   \label{NtoMdiag}
\end{align}
having defined $P_+^{M \otimes N}  :  =  P_+^{(M)}   \otimes P_+^{(N)} $ and
\begin{align*} 
      |\Phi_n\>  :  =    P_{+}^{M\otimes
N}     \left(      |I^{\otimes M}  \>\!\>  \otimes
  |\varphi_n\>  \right) \qquad  |I^{\otimes M}  \>\!\>  : =\bigotimes_{i=1}^{M}
|I\rangle\!\rangle_{i i} . 
\end{align*}
Eq. (\ref{NtoMdiag})  is the desired diagonalization of the Choi operator
$C^+_{q ,N \to M}$: Indeed, we have 
\begin{align*}
\<\Phi_{m}  | \Phi_{n}\>  &  =     \< \!\<  I^{\otimes M}|    \< \varphi_m|    
\left (P^{(M)}_+  \otimes P^{(N)}_{+}   \right)    |I^{\otimes M}\>\!\>
|\varphi_n\>   \\
  &  =   \<  \varphi_m|     \Tr_{1, \dots, M}  \left [P^{(N)}_{+}  \right]    
|\varphi_n\>   \\
  &  = \frac{d^{(N)}_+}{d^{(N-M)}_+}   ~     \<  \varphi_m|      P^{(N-M)}_{+}  
 |\varphi_n\>  \\
   & =  \frac{d^{(N)}_+}{d^{(N-1)}_+}  ~ \delta_{mn}  , 
\end{align*}
that is, the vectors $\{|\Phi_n \>\}$ are mutually orthogonal and  have the
same norm.  Therefore, $C^+_{q, N \to M}$ only has one non-zero eigenvalue,
given by  $\gamma_{N\to M} = \frac{d^{(N)}_+}{d^{(N-M)}_+}$.     

The causality bound of Eq. (\ref{boundNM}) is equivalent to $p  ~  \gamma_{N \to
M}  \le    \frac 1 {d^{(M)}_+}$, which implies that the maximum probability is
given by 
\begin{align}
p_{q, N \to M}  =  \frac{d^{(N-M)}}{d^{(N)}_+  d^{(M)}_+}.
\end{align}

Note that $p_{q, N \to M}$ converges to the value
\begin{align*}
p_{q, \infty \to M} =  \frac 1 {d^{(M)}_+}
\end{align*}
in the limit of $N$ going to infinity.  Such an asymptotic behaviour will be
explained in Section \ref{sec:summary}.

\subsection{From $N$ copies of a pure state in the future to $M >N$ optimal
clones in the past}

We now generalize the results of the previous subsection to the case $M>N$.   Of
course, in this case we cannot
expect to obtain $M$ perfect output copies of the input state,  because that
would violate the no-cloning
theorem\cite{no-cloning,no-cloning2}.   What we can do is to consider the best
possible approximation of the ideal cloning process.  Such an optimal
approximation is given by the optimal \emph{universal cloning} channel  
\cite{cloning},  here denoted by $\map C_{q, N\to M}^+$ for consistency of
notation.  The cloning channel $\map C_{q, N\to M}^+$  is given by 
\begin{align}\label{clonNM}
\mathcal{C}^+_{q, N \to M}(\rho)=\frac{d_{+}^{(N)}}{d_{+}^{(M)}}P_{+}^{(M)} 
\left (  P^{(N)}_+  \rho  P^{(N)}_+  \otimes
I^{\otimes (M-N)}  \right)  P_{+}^{(M)}.
\end{align}
Note that the linear map $\map C_{q, N \to M}^+$  is trace-preserving only for
density matrices $\rho$ with support within the symmetric subspace $\left(  
\hilb^{\otimes N} \right)_+$.  Indeed,  $\map C_{q, N \to M}^+$ should be
regarded as a quantum channel sending states on $\hilb_{in}  :  = \left(  
\hilb^{\otimes N} \right)_+$ to states on $\hilb_{out} :  = \left(  
\hilb^{\otimes M} \right)_+$.   

From Eqs.  (\ref{CNM})  and  (\ref{clonNM}),   we can note the existence of an
important duality between universal cloning and partial trace:  for two
arbitrary operators $\nu$  on  $\left(   \hilb^{\otimes N} \right)_+$  and $\mu
$  on    $\left(   \hilb^{\otimes M} \right)_+$ one has 
\begin{align}
   \Tr[  \mu    ~ \map C_{q,  N \to M}^+     (\nu)   ]  =    \frac{ d_+^{(M)}
}{d_+^{(N)}}  ~   \Tr  [   \map C^+_{q, M \to N}    (  \mu)  ~  \nu   ].  
\end{align} 
In other words, the  map representing   the universal cloning from $N$ to $M$
copies in the Heisenberg picture    is the map representing the partial trace
from $M$ to $N$ copies in the Schr\"odinger picture, up to a positive scaling
factor.  We will refer to this property as the \emph{time-reversal duality}
between partial trace and universal cloning.


In terms of Choi operators, the duality can be written as 
\begin{align}\label{clontracechoiduality}
C_{q, N\to M}^+     =   \frac{d_+^{(N)}}{d_+^{(M)}} ~    {S}_{N,  M}     C_{q,
M\to N}^{+}   {S}_{N,  M} ,
\end{align}
where $S_{N,M}$  is the operator that exchanges the $N$ output Hilbert spaces
with the $M$ input Hilbert spaces for the channel $\map C^{+}_{q, M \to N}$.  

Eq. (\ref{clontracechoiduality})  allows us to easily find the probability of
success for the cloning channel $\map C_{q, N \to M}^+$. 
First of all,   Eq. (\ref{clontracechoiduality})   implies that  $C_{q, N\to
M}^+  $  has the symmetry    $(U^{\otimes M}  \otimes U^{*\otimes N})  C_{q,
N\to M}^+  (U^{\otimes M}  \otimes U^{*\otimes N})  =  C_{q, N\to M}^+  $  for
every $U  \in  SU(d)$.  Using this symmetry the causality bound becomes 
\begin{align*}
p ~   C_{q, N\to M}^+    \le  \frac{P_+^{(M)}  \otimes P_+^{(N)}}{d_+^{(M)}}    
\end{align*}
Now, Eq. (\ref{clontracechoiduality}) implies that the above equation is
equivalent to 
\begin{align*}
p ~   C_{q, M\to N}^+    \le  \frac{P_+^{(N)}  \otimes P_+^{(M)}}{d_+^{(N)}},  
\end{align*}
which is nothing but the causality bound for the trace channel $\map C_{q,  M
\to N }^+$  [cf. Eq. (\ref{boundNM})].    Therefore, we have that the causality
bound for the cloning channel  $\map C_{q,  N \to M }^+$ is satisfied with
probability $p$ if and only if the causality bound for the trace
channel $\map C_{q,  M \to N }^+$ is satisfied with probability $ p $.    Hence,
we have 
\begin{align*}
p^+_{q, N \to M}  =  p^+_{q, M \to N}  =  \frac{d_+^{(M-N)}}{d_+^{(M)} 
d_+^{(N)}}.
\end{align*}
Quite surprisingly, the above value increases with the number $M$ of output
copies, reaching the asymptotic value 
\begin{align*}
p_{q, M \to \infty}  =  \frac 1 {d_+^{(N)}}  
\end{align*}
in the limit $M \to \infty$.  This asymptotic behaviour will be explained in
Section \ref{sec:summary}.

\section{Dependence of the  probability of success on the amount and type of
information transmitted by the channel}\label{sec:summary}

The results obtained so far indicate a dependence of the maximum probability of
success  on the amount and type (quantum or classical) of information that the
channel can transmit:    First, we noted that the only channels that can be
simulated with unit probability are the erasure channels, that produce a fixed
output state independently of the input,---that is, they do not transmit any
information.  Then, we saw that measure-and-prepare channels, that only transmit
classical information, can be simulated with probability $p_{m \& p}  \ge
1/d_{out}$, $d_{out}$  being the dimension of the output Hilbert space.    
Finally, channels that transmit both quantum and classical information can be
simulated with probability  $p \ge   \max  \{  1/d^2_{in},   1/d^2_{out} \}$, as
in Eq. (\ref{generallower})     

In the following we will make a quantitative connection between the amount of
information transmitted by a channel and the probability of simulating it with a
generalized teleportation protocol: we will show that the probability of success
is upper bounded by the inverse of a measure of information defined as the
maximum payoff in a communication game.  We will then use this result to explain
the asymptotic behaviour of the probability of success for the simulation of a
partial
trace channel with a large number of input systems and of a universal cloning
channel with
a large number of output systems.

\subsection{Statistical information  bounds}
In this subsection we relate probability of success in the simulation of a
channel to the past with the amount of information transmitted by the channel.
In this context, a convenient measure of information is the expected payoff
obtained by two parties in a communication game.    

Suppose that the channel $\map C$ is used to transfer states from Alice's side
to states on Bob's side.  
Alice can encode a set of messages $X$ in a set of states $\{  \rho_i    \}_{i
\in X} $ on $ \hilb_{in}$ and  communicate a message to Bob by sending the
corresponding state through the channel $\map C$.      On his side,   Bob can
try to decode Alice's message from the output state $\map C(\rho_i)$. 
Generally, the decoding will be described by a POVM  $\{P_j\}_{j\in Y}$ on the
output Hilbert space $\hilb_{out}$, with the outcome $y$  corresponding to the
decoded message.    Usually, but not necessarily, the set of decoded messages
$X$ coincides with the set of encoded messages $Y$.        To assess the success
of the communication, we can introduce a payoff function $f:  X \times Y  \to
\mathbb R$, that quantifies the gain (or the loss) associated to decoding $j$
while the encoded message was $i$.   For given $i$, the expected payoff will be
\begin{align*}
\mathbb E_i  (  f)   :=   \sum_{j \in Y}           f(i,j)~   \Tr[ \map C(\rho_i)
   P_j ].  
\end{align*} 

Now, suppose that Alice is in the future and Bob is in the past, and that they
are using a probabilistic simulation of channel $\map C$.  The causality bound
of Eq. (\ref{MapsInequality})  imposes $ p~    \map C(\rho_i)  \le  \rho_0  $,
and, therefore
\begin{align}\label{fidboundi}
p ~  \mathbb E_i  (  f)     \le  \mathbb  E_{i,\rho_0}  (f)  \qquad   \forall i
\in X   , 
\end{align}  
  having defined   $\mathbb  E_{i,\rho_0}    :  =  \sum_{j \in Y}          
f(i,j)~   \Tr[  \rho_0  P_j ]$. 
  Introducing a prior probability distribution $\{\pi_i\}_{i \in X}$ on Alice's
messages we can then
consider the average expected payoff
  \begin{align*}
  \mathbb E_{ave} (f) :  =   \sum_{i \in  X }  \pi_i    ~ \mathbb E_i  (  f)  
  \end{align*}
Hence, the bound of Eq. (\ref{fidboundi}) becomes  
\begin{align*}
p ~  \mathbb E_{ave}  (  f)     &\le  \sum_{i\in X}   \pi_i   ~  \mathbb 
E_{i,\rho_0} (f)\\
  &  \le        \max_{j  \in  Y}    \left( \sum_{i\in X}  \pi_i ~ f(i, j)
\right)  :  =  f_\pi.
\end{align*} 
Note that the value $f_\pi$ depends only on the payoff function $f$ and on the
prior probabilities $\{\pi_i\}_{i \in X}$.

We have now arrived at the bound 
\begin{align}\label{infobound}
p_{\map C}   \le  \frac{f_\pi}{  \mathbb E_{ave} (f)  }, 
\end{align}
which holds for every given payoff function $f$,  prior probabilites
$\{\pi_i\}$, encoding $\{  \rho_i \}_{i \in  X}$, and decoding $\{ P_j \}_{j \in
Y}$.
Maximizing the average payoff over all possible choices of encoding $\{  \rho_i
\}_{i \in  X}$ and decoding $\{ P_j \}_{j \in Y}$ we then get the maximum  value
 
\begin{align*}
\mathbb E_{\max} (f)  :=  \max_{\{ P_j \}_{j \in Y}  }  \max_{\{ \rho_i\}_{i \in
 X}}     \mathbb E_{ave}  (  f) ,
\end{align*}
which quantifies the maximum payoff that can be achieved when using the channel
$\map C$ in a communication game with given payoff function $f$ and  prior
probabilities $\{\pi_{i}\}_{i \in X }$.   

We then conclude that the maximum probability to simulate the channel $\map C$,
denoted by  $p_{\map C}$, has to satisfy the bound 
\begin{align*}
p_{\map C}   \le  \frac{f_\pi}{  \mathbb E_{\max} (f)  },
\end{align*}  
for every payoff function $f$ and for every prior  $\{\pi_{i}\}_{i \in X }$. The
above bound states that the probability of success cannot be greater than the
inverse of the maximum amount of statistical information (average payoff)
transmitted by the channel. For this reason, we refer to this bound as the
\emph{statistical information bound}.


Notice that the derivation of the information bound in Eq. (\ref{infobound}) can
be extended to the case where Alice and Bob are allowed to communicate through
an ideal quantum channel on a reference system with Hilbert space $\hilb_R$, in
addition to the channel $\map C$.  In
this case, Alice can encode her message $i\in  X$  in a bipartite state
$\rho_{i}$ on $\hilb_{in}  \otimes \hilb_R$. Later, she will send the input
system through the channel $\map C$ and the reference through the ideal channel.
 
Bob's task will then be to decode the message from the output state $(\map C 
\otimes   \map I_R)  (\rho_i)$, through a bipartite POVM $\{P_j\}_{j \in Y}$  on
 $\hilb_{out} \otimes \hilb_R$.     Following the same steps as above, in this
case we obtain the \emph{entanglement-assisted statistical information bound}  
\begin{align*}
p_{\map C}  \le  \frac{\mathbb E^{(R)}_{ave}  (f) } {  \mathbb E^{(SR)}_{ave} 
(f) },
\end{align*}   
having defined
\begin{align*}
 \mathbb E^{(SR)}_{ave}  (f)    : =  \sum_{i\in X}  \pi_i   \left\{   
\sum_{j\in Y}   f(i,j)  ~  \Tr\left[ (\map C  \otimes   \map I_R)  (\rho_i)   
P_j \right]  \right\}\\
  \mathbb E^{(R)}_{ave}  (f)    : =  \sum_{i\in X}  \pi_i   \left\{   
\sum_{j\in Y}   f(i,j)  ~  \Tr\left[ \rho_0  \otimes \Tr_{in}[\rho_i] )   P_j
\right]  \right\}.
\end{align*}

Note, however, that both statistical information bounds are just upper bounds:
In general  there is no guarantee about their achievability.      

\subsection{Asymptotic behaviour of the optimal probabilistic teleportation from
$N$ to $M \le N$ copies}  
The aim of the following discussion is to provide an explanation for the
asymptotic behaviour   
  of the probability of success in the simulation of  the partial trace channel
$\map
C^+_{q, N \to M}$ in the limit $N \to \infty$.  
First, we apply the statistical information bound of Eq. (\ref{infobound}) to
show that the probability of success satisfies the bound 
\begin{align*}
p^+_{q, N \to M}  \le \frac{1}{d_+^{(M)}}  \qquad \forall  N \ge M. 
\end{align*}
Consider a communication scenario where Alice encodes a message in  state  from
the set $\{  |\varphi\>^{\otimes N}\}_{|\varphi\> \in \hilb}$ and sends it to
Bob through the partial trace channel $\map C_{q, N \to M}$. Suppose that  Bob's
decoding is given by the  POVM 
\begin{align*} 
 P^{(M)}_\psi  ~ d \psi  = d_{+}^{(M)}   ~  \left(  |\psi\>\<\psi| 
\right)^{\otimes M}   d \psi ,
 \end{align*} 
 where $d \psi$ is the normalized invariant measure on the pure states of
$\hilb$.    As a payoff function, let us assume the Dirac delta $f (  \varphi,
\psi)  =  \delta(\varphi-\psi)$, so that the expected payoff for input
$|\varphi\>$ is just the probability (density) to make the correct guess
\begin{align*}  
\mathbb E_{\varphi}  (f)  &=  \int  d \psi   ~  \delta(\varphi-\psi) ~ 
\Tr\left\{   \map C^+_{q, N \to M} \left[   \left (|\varphi\>\<\varphi| 
\right)^{\otimes N} \right] P^{(M)}_\psi \right\} \\
  & =\Tr[  \left (|\varphi\>\<\varphi|  \right)^{\otimes M}    P^{(M)}_\varphi ]
 \\
  &  =  d_+^{(M)}  \qquad \forall   |\varphi\>  \in  \hilb.  
\end{align*}  
Since the expected payoff is the same for every state, we have 
\begin{align*}
\mathbb E_{ave}  (f)  =  d_+^{(M)}
\end{align*}   
for every prior distribution.
On the other hand, choosing the uniform distribution $\pi(d \varphi)  =  d
\varphi$  for the input states,  we have 
\begin{align*}
f_\pi   & =    \max_{\psi}  \left[ \int  d \varphi   ~  \delta(\varphi-\psi)
\right]  \\
  &  =  1.     
\end{align*}
The statistical information bound of Eq. (\ref{infobound}) then gives  $p^+_{q,
N \to M}  \le \frac 1 {d_+^{(M)}}$.  

In subsection  \ref{subsect:traceNM}  we observed that the probability of
success $p^+_{q, N \to M}$ asymptotically approaches the value $p^+_{q, \infty
\to M}  = 1/ {d_+^{(M)}}$ in the limit $N \to \infty$.  
  An explanation of this behaviour is the following:  From Ref.
\cite{definetti}, we know that in the limit $N \to \infty$ the partial trace
channel $\map C^+_{q, N \to M}$ converges to the estimation channel  $ \map
C^+_{est, N \to M}  
$ defined by
  \begin{align*}
  \map C^+_{est, N \to M} (\rho):  = \int  d \psi     \left(  |\psi\>\<\psi| 
\right)^{\otimes M}   \Tr[  \rho  P^{(N)}_\psi]  . 
  \end{align*}
 with $P^{(N)}_\psi  :  =  d_+^{(N)}      \left(  |\psi\>\<\psi| 
\right)^{\otimes N}$. 
 Hence, the probability of success $p^+_{q, N \to M}$ must converge to
$p^+_{est, N \to M}$, the probability of success for the estimation channel  $ 
\map C^+_{est, N \to M}  
$.  On the other, hand, since  $ \map C^+_{est, N \to M}  
$ is a measure-and-prepare channel, by the general bound of Eq. (\ref{mpbound})
we must have $p^+_{est, N \to M}  \ge  1/d_+^{(M)}$.  In conclusion, we
obtained 
\begin{align*}
\frac 1  {d_+^{(M)}}   \ge  \lim_{N \to \infty}  p^+_{q, N \to M}  =\lim_{N \to
\infty} p^+_{est, N \to M}     \ge      \frac 1  {d_+^{(M)}}  ,  
\end{align*}
that is, $\lim_{N \to \infty}  p^+_{q, N \to M}  =  \frac 1  {d_+^{(M)}} $.  In
other words, the asymptotic behaviour of  $p^+_{q, N \to M}$ is dictated by the
statistical information bound and by the
convergence of the trace channel $\map C_{q,N\to M}^+$ to a measure-and-prepare
channel.

\begin{figure}[h]
\begin{centering}
\includegraphics[scale=0.5]{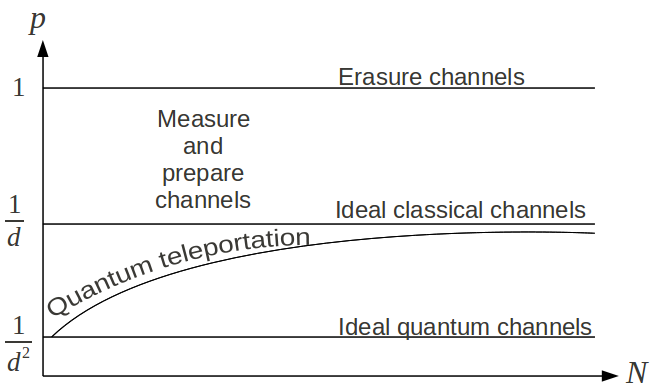}
\par\end{centering}

\caption{\label{fig:channelsGraph}A graph of the maximum probability $p$ to
simulate the partial trace channel from $N$ copies to $M=1$ copy of a pure input
state.
The maximum probabilities for erasure channels, measure-and-prepare channels,
and ideal classical and quantum channels are also located in the plot.   }
\end{figure}

 An illustration of the situation for the case of $M=1$  is given in Fig.
\ref{fig:channelsGraph}. It is worth stressing, once more,  that no improvement
with the number of input copies would be possible  for a classical partial trace
channel: Since in that case  the probability is already $p_{cl}  =\frac{1}{d}$,
any improvement would violate the causality bound for ideal classical channels.
Hence, the increase of the probability of teleportation with the number $N$ of
input copies is intrinsically a  quantum phenomenon.

\subsection{Asymptotic behaviour of the optimal probabilistic teleportation from
$N$ to $M \ge N$ copies}    

Here  we consider  the universal cloning channel $\map C^+_{q, N \to M}$. As we
did for the trace channels,  we will now provide an explanation for the
asymptotic behaviour of the probability of success in the limit $M \to \infty$.

A direct explanation comes from the time-reversal duality between cloning and
partial trace, which implies the relation $p^+_{q, N \to M}  =  p_{q, M \to
N}^+$ and thus allows one to reduce the asymptotic behaviour of the probability
of success of a cloning channel with many output copies to the asymptotic
behaviour of the probability of success for a trace channel with many input
copies.  

A less direct, but perhaps more insightful explanation  can be obtained by
following  the same steps used in the previous subsection.
  Let us start by showing that  the statistical information bound of Eq.
(\ref{infobound}) implies that the probability of success satisfies the bound 
\begin{align*}
p^+_{q, N \to M}  \le \frac{1}{d_+^{(N)}}  \qquad \forall  M \ge N. 
\end{align*}
Consider a communication scenario where Alice encodes a message in  state  from
the set $\{  |\varphi\>^{\otimes N}\}_{|\varphi\> \in \hilb}$ and sends it to
Bob through the cloning channel $\map C_{q, N \to M}$. Suppose that  Bob's
decoding is given by the  POVM $  P^{(M)}_\psi  ~ d \psi  = d_{+}^{(M)}   ~ 
\left(  |\psi\>\<\psi|  \right)^{\otimes M}   d \psi $, where $d \psi$ is the
normalized invariant measure on the pure states of $\hilb$.    As a payoff
function we assume again the Dirac delta $f (  \varphi, \psi)  = 
\delta(\varphi-\psi)$, so that the expected payoff for input $|\varphi\>$ is the
probability (density) of the correct guess:
\begin{align*}  
\mathbb E_{\varphi}  (f)  &=  \int  d \psi   ~  \delta(\varphi-\psi) ~ 
\Tr\left\{   \map C_{q, N \to M}^+  \left[ \left (|\varphi\>\<\varphi| 
\right)^{\otimes N}  \right]    P^{(M)}_\psi \right\}  \\
  & =    
   \Tr\left\{   \map C_{q, N \to M}^+  \left[ \left (|\varphi\>\<\varphi| 
\right)^{\otimes N}  \right]    P^{(M)}_\varphi \right\}     \\
    & =     \frac {d_+^{(N)}}  {d_+^{(M)}} 
   \Tr\left [  \left (|\varphi\>\<\varphi|  \right)^{\otimes N}  \map C^+_{q, M
\to N}    \left(   P^{(M)}_\varphi  \right) \right]    \\
   & = \Tr[   \left (|\varphi\>\<\varphi|  \right)^{\otimes N}   
P^{(N)}_\varphi ]  \\
  &  =  d_+^{(N)}  \qquad \forall   |\varphi\>  \in  \hilb.  
\end{align*}  
Since the expected payoff is the same for every state, for every prior
distribution we have $\mathbb E_{ave}  (f)  =  d_+^{(N)}$.   
On the other hand, we already know that $f_\pi  =  1$, so that the  information
bound of Eq. (\ref{infobound}) gives  $p_{q, N \to M}  \le \frac 1 {d_+^{(N)}}$.

On the other hand, from Ref. \cite{definetti} we know that in the limit $M \to
\infty$ the cloning channel $\map C^+_{q, N \to M}$ converges to the estimation
channel  $ \map C^+_{est, N \to M}  
$.  Hence, the probability of success $p^+_{q, N \to M}$ must converge to
$p^+_{est, N \to M}$. Again, we obtain the relation
\begin{align*}
\frac 1  {d_+^{(N)}}   \ge  \lim_{M \to \infty}  p^+_{q, N \to M}  =\lim_{M \to
\infty} p^+_{est, N \to M}     \ge      \frac 1  {d_+^{(N)}}  ,  
\end{align*}
that is, $\lim_{M \to \infty}  p^+_{q, N \to M}  =  \frac 1  {d_+^{(N)}} $. 
Also in this case, the asymptotic behaviour of the probability $p^+_{q, N \to
M}$ is dictated by the statistical information bound and by the convergence of
the cloning channel $\map C_{q, N \to
M}^+$ to a measure-and-prepare channel.

\section{Conclusions}\label{sec:conclusion}
This work showed how causality determines the maximum probability of simulating
a
given quantum channel from the future to the past.  
Since causality states that  it is not possible to signal from the future to the
past,  the probability of success in the simulation should be small enough to
prevent any signal from being sent.     As
a consequence, channels that are able to transmit more information will have a
smaller probability of being simulated, while channels that transmit less
information will have a larger probability.   In particular, erasure channels,
that transmit no information at all, can be simulated with unit probability,
while measure-and-prepare channels, that transmit only classical information,
can be simulated with probability at least equal to $p_{cl}  =  1/d_{out}$  the
inverse of the output Hilbert space dimension.  The hardest channels to 
simulate are the ideal quantum channels, for which the probability is equal to
$p_q  =  1/d^2$,  the inverse of the square of the Hilbert space dimension.  A
quantitative connection between the amount of information transmitted by a
channel and the probability of successful simulation is given by the statistical
information bounds, which state that the probability of success is upper bounded
by the inverse of the statistical information transmitted by the channel.

In the paper we computed explicitly the maximum probability for the simulation
of several channels, including partial trace channels from $N$ input systems to
$M \le
N$  output systems, and universal cloning channels from $N$ input systems to $M
\ge N$ output systems.  We pointed out a time-reversal duality between trace and
cloning and we exploited it to show  that the value of the maximum probability
of
success is given by  $p^+_{q, N \to M}  =  d_+^{(|N-M|}/  \left(  d_+^{(M)}
d_+^{(N)}\right) $.   
Note that for fixed $M$, the probability of success is an increasing
function of $N$, the number of input copies.  Similarly, for fixed $N$, the
probability of success is an increasing function of $M$, the number of output
copies.  These are  genuinely quantum features, that are impossible in the
classical world:  Indeed, classically the probability of success for all trace
channels and for all cloning channels is given by $p_{cl}  = 1/d $,
independently of $N$ and $M$. 

In the case of a single output copy, $M  =1$ the probability of success for a
quantum partial trace channel tends to the classical value $p_{cl}  = 1/d $ in
the limit of $N$ going to infinity.  
This asymptotic behaviour can be explained from the fact that a trace channel
with asymptotically large number of input copies converges to a
measure-and-prepare channel \cite{definetti}.   The same explanation can be
provided for the
asymptotic behaviour of the probability of success for cloning channels. Also in
this case the asymptotic behaviour is dictated by the fact that a universal
cloning channel with asymptotically large number of output copies converges to a
measure-and-prepare channel  \cite{definetti}.

Note that in this paper we focused on determining the maximum probability of
success for many channels, but we did not show explicitly the generalized
teleportation schemes that achieve the simulation of those channels.  In
particular, it is interesting to ask which kind of entangled states and which
kind of measurements  allow one to achieve teleportation from $N$ input copies
to $M$ output copies. Moreover, it is worth asking whether these probabilistic
teleportation protocols can be extended to deterministic protocols  (now, from
the past to the future) by reintroducing classical communication and correction
operations, as in the case of tele-cloning \cite{teleclon}.   The fact that the
probability of successful teleportation from $N$ copies  to $M=1$ copy converges
to $p_{cl} =  1/d $ suggests that our probabilistic protocol could be extended
to a deterministic protocol that
needs only $\log(d)$ bits of classical communication, instead of the $2\log(d)$
bits of the original teleportation protocol.  However, it is quite possible that
in general there exists a trade-off between maximizing the probability of
success for a particular outcome and being able to implement a correction for
all outcomes.  All these issues will be investigated in a forthcoming work
\cite{inprep}.

Another interesting direction of future research is to consider the simulation
of multipartite channels under the constraint that different input and output
systems can arrive at different instants of time, and to show how the
probability of simulation depends on the possible causal arrangements of the
input/output systems.  For example, one could consider the case of a cloning
channel from $N=1$ input copy to $M=2$ approximate copies, one copy being sent
to the past and  the other being sent to the future.  Examples of this kind are
expected to shed light on the interplay between causal structure and quantum
information flow. 

\acknowledgments 

Research at Perimeter Institute for Theoretical Physics is supported in part by
the Government of Canada through NSERC and by the Province of Ontario through
MRI.

\end{document}